\documentclass[11pt]{article}
\usepackage{fullpage}
\usepackage{cite}
\usepackage{graphicx}
\usepackage{amsmath}
\usepackage{amssymb}
\title{}
\date{}
\renewcommand{\vec}[1]{\mbox{\boldmath$ #1 $}}

\begin{document}
\bibliographystyle{utphys}
\newcommand{\msbar}{\ensuremath{\overline{\text{MS}}}}
\newcommand{\DIS}{\ensuremath{\text{DIS}}}
\newcommand{\abar}{\ensuremath{\bar{\alpha}_S}}
\newcommand{\bb}{\ensuremath{\bar{\beta}_0}}
\newcommand{\rc}{\ensuremath{r_{\text{cut}}}}
\newcommand{\Nd}{\ensuremath{N_{\text{d.o.f.}}}}
\setlength{\parindent}{0pt}
\def\kh{\hat{k}}

\titlepage
\begin{flushright}
Edinburgh 2014/18 \\
\end{flushright}

\vspace*{0.5cm}

\begin{center}
{\Large \bf Black holes and the double copy}
\end{center}

\vspace*{1cm}

\begin{center} 
\textsc{R. Monteiro$^a$\footnote{monteiro@maths.ox.ac.uk},
D. O'Connell$^b$\footnote{donal@staffmail.ed.ac.uk},
C. D. White$^c$\footnote{Christopher.White@glasgow.ac.uk} } \\

\vspace*{0.5cm} $^a$ Mathematical Institute, University of Oxford, \\Oxford OX2 6GG, England, UK\\

\vspace*{0.5cm} $^b$ Higgs Centre for Theoretical Physics, \\School of Physics and Astronomy, The University of Edinburgh,\\
Edinburgh EH9 3JZ, Scotland, UK\\

\vspace*{0.5cm} $^c$ SUPA, School of Physics and Astronomy, University of Glasgow,\\ Glasgow G12 8QQ, Scotland, UK\\

\end{center}

\vspace*{0.5cm}

\begin{abstract}
Recently, a perturbative duality between gauge and gravity theories
(the double copy) has been discovered, that is believed to hold to all
loop orders. In this paper, we examine the relationship between
classical solutions of non-Abelian gauge theory and gravity. We
propose a general class of gauge theory solutions that double copy to
gravity, namely those involving stationary Kerr-Schild metrics. The
Schwarzschild and Kerr black holes (plus their higher-dimensional
equivalents) emerge as special cases. We also discuss plane wave solutions. Furthermore, a recently examined
double copy between the self-dual sectors of Yang-Mills theory and
gravity can be reinterpreted using a momentum-space generalisation of
the Kerr-Schild framework.
\end{abstract}

\vspace*{0.5cm}

\section{Introduction}
There is a growing body of work examining the relationships between
(non)-Abelian gauge theories and gravity, both with and without
supersymmetry. Our primary interest in this article will be the {\it
  double copy} of
ref.~\cite{Bern:2008qj,Bern:2010yg,Bern:2010ue}. This postulates that
scattering amplitudes in non-Abelian gauge theories can be expressed
such that replacement of colour information by additional kinematic
dependence, in a well-defined way, automatically leads to gravity
amplitudes. For this to work, a certain duality between colour and
kinematics -- {\it BCJ duality} -- has to be made manifest in the
gauge theory. \\

BCJ duality, and the double copy, have been proven at tree
level~\cite{BjerrumBohr:2009rd,Stieberger:2009hq,Bern:2010yg,BjerrumBohr:2010zs,Feng:2010my,Tye:2010dd,Mafra:2011kj,Cachazo:2012uq,BjerrumBohr:2012mg}. The
latter is then equivalent to the well-known KLT
relations~\cite{Kawai:1985xq}, whose origin is the relationship
between open and closed string tree amplitudes. Remarkably, however,
both BCJ duality and the double copy appear to be true at loop level,
including in the absence of
supersymmetry~\cite{Bern:2010ue,Bern:1998ug,Green:1982sw,Bern:1997nh,Oxburgh:2012zr,Carrasco:2011mn,Carrasco:2012ca,Bargheer:2012gv,Boels:2013bi,Bjerrum-Bohr:2013iza,Bern:2013yya,Bern:2013qca,Nohle:2013bfa,Bern:2013uka,Naculich:2013xa,Du:2014uua,Bern:2014sna}. Other
investigations have verified their application in form factors rather
than scattering amplitudes~\cite{Boels:2012ew}, and also in theories
with fundamental matter~\cite{Johansson:2014zca}. All-order tests have been
performed in the soft limit of fixed-angle
scattering~\cite{Oxburgh:2012zr}, and the Regge
limit~\cite{Saotome:2012vy,Vera:2012ds,Johansson:2013nsa,Johansson:2013aca}
(see
also~\cite{Fu:2014pya,Bianchi:2013pfa,Monteiro:2013rya,Tolotti:2013caa,Fu:2013qna,Du:2013sha,Fu:2012uy,Sivaramakrishnan:2014bpa,Naculich:2014naa,Naculich:2014rta,Chiodaroli:2014xia,Carrasco:2013ypa,Litsey:2013jfa}
for related studies). Despite this progress, however, it is perhaps
fair to say that we do not fully understand the implications of BCJ
duality, and the associated double copy. The main reason for this is
that both ideas are currently defined in a purely perturbative
context. Consequently, exploration of the deeper explanation of BCJ
duality is hampered by the complexity of multiloop calculations.
Motivated by this, one may
ask whether any features of the double copy manifest themselves in a
classical
context. If the answer
to this question is yes, this potentially offers a significant insight
into gauge and gravity theories.\\

In fact, a number of cases where classical solutions match up between
gauge and gravity theories already exist in the literature. The case
of linearised waves is quite obvious; see e.g.~\cite{Siegel:1999ew}. A
more involved example is that of (non)-Abelian shockwaves, recently
studied in \cite{Saotome:2012vy} in the context of the high-energy
(Regge) limit, where Feynman diagrams can be summed to all orders,
exhibiting a double copy between the two theories. One attempt at a
Lagrangian-level understanding of the double copy arises after
projection to the self-dual sector of both theories, as examined in
ref.~\cite{Monteiro:2011pc}. There the action can be made manifestly
cubic~\cite{Parkes:1992rz} in terms of an adjoint-valued scalar
field. The 3-vertex for this field has the form of a product of two
structure constants, one for the gauge group, and the other for a
kinematic Lie algebra of area-preserving diffeomorphisms. Thus, BCJ
duality is made manifest, and the double copy relationship to
self-dual gravity is also straightforward.\\

Another framework for understanding the double copy at the Lagrangian
level has been presented in ref.~\cite{Bern:2010yg} (see also
\cite{Tolotti:2013caa} and the more recent light-like gauge approach \cite{Vaman:2014iwa}). In that paper, additional terms are added to
the Lagrangian of Yang-Mills theory order by order in perturbation
theory, such that Feynman diagrams calculated with this effective
Lagrangian are manifestly BCJ-dual. The double copy of this gives a
gravitational Lagrangian as expected. However, it is not yet known how
to generalise this construction to the nonperturbative level, in that
a closed form for the all-order Lagrangian is not
available. Beyond Lagrangian approaches, the recently
studied scattering equations provide formulas for gauge theory and
gravity amplitudes which exhibit an alternative form of double copy
\cite{Cachazo:2013hca,Cachazo:2013iea,Cachazo:2014nsa}. Those formulas
can be directly obtained from the new ambitwistor string theories,
where the distinction between gauge theory and gravity is expressed by
a natural choice of worldsheet fields
\cite{Mason:2013sva,Geyer:2014fka}. Closer to our concerns here, the
recent work \cite{Anastasiou:2014qba} considers the relation between
gauge symmetries and gravitational symmetries at the linearised level
using a double copy prescription based on the convolution of
fields. \\

The aim of this paper is to find more examples of classical solutions
which match under the double copy.  We observe that Kerr-Schild
coordinates in gravity provide a natural relation to solutions in gauge theory. By examining this connection, we find an
infinite class of solutions with the double-copy property.  In
particular, we will find single copies, or ``square roots'', of the
Schwarzschild\footnote{A previous proposal for a single copy for the
  Schwarzschild black hole has been made by relating the field
  strength and curvature tensors in the gauge / gravity theories. Our
  proposal differs from this as it is set up directly in terms of the
  gluon and graviton fields directly. We thank Nima Arkani-Hamed for
  discussions on this point, and Clifford Cheung for providing
  ref.~\cite{Cheungnotes}. } and Kerr black holes (plus their higher
dimensional
generalisations). The proposed relation between the Schwarzschild and Coulomb solutions, for instance, is consistent with the expectation from the construction of these solutions in perturbation theory; see e.g.~\cite{Neill:2013wsa}. \\

An interesting aspect of our work is the importance of a ``zeroth"
copy of the gauge theory result.  This consists of replacing kinematic
information in the gauge theory with additional colour information,
and leads to a {\it biadjoint scalar theory} i.e. a cubic scalar
theory, in which the scalar transforms in the adjoint of two Lie
groups. Such theories have been shown to be increasingly important in
studies of BCJ duality and the double
copy~\cite{Bern:1999bx,Du:2011js,BjerrumBohr:2012mg,Cachazo:2013iea,Anastasiou:2014qba}. The
biadjoint scalar theory will also provide a physical interpretation of
the double copy procedure, that ties it to the usual formulation
involving perturbative scattering amplitudes.\\

As we will review in what follows, the major ingredients in the double
copy story (provided BCJ duality is satisfied) are:
\begin{itemize}
\item {\it Colour factors}. These are replaced by kinematic
  numerators, which has the effect of replacing Lorentz vectors by an
  outer product of two Lorentz vectors.
\item {\it Scalar propagators}. Denominators of individual cubic
  diagrams are not replaced in the double copy procedure. These
  denominator factors can be interpreted as scalar propagators
  associated with the topology of the diagram.
\end{itemize}
In our examples, we will encounter clear analogues of these
ingredients. Each example has a Lorentz vector in the gauge theory,
such that the corresponding gravity solution consists of an outer
product of this vector with itself. Scalar prefactors also arise,
fixed by the zeroth copy, that are not squared when performing the
double copy procedure. Furthermore, we will see explicitly that these
scalar functions can be physically interpreted as scalar propagators
integrated over source distributions of charge / mass. Whilst we
remain far from understanding the general circumstances in which
classical solutions must match up between Yang-Mills theory and
gravity (including the appropriate gauges to be chosen on either side
of the correspondence), we hope that our results provide a useful
springboard for a more systematic and far-reaching investigation. \\

The structure of the paper is as follows. In
section~\ref{sec:amplitude}, we briefly review aspects of BCJ duality
and the double / zeroth copies. In section~\ref{sec:Kerr-Schild}, we
review properties of Kerr-Schild coordinates, and show how the
self-dual double copy of ref.~\cite{Monteiro:2011pc} can be cast in
this language. In sections~\ref{sec:Kerr2}, we will examine familiar
examples of GR solutions that emerge as special cases of the
Kerr-Schild framework. In section~\ref{sec:waves}, we consider some
time-dependent solutions, which correspond to non-perturbative waves. Finally, we discuss our
results and conclude in section~\ref{sec:conclusion}.

\section{The double copy}
\label{sec:amplitude}

In this section, we will review the well-known double copy story for
amplitudes~\cite{Bern:2008qj,Bern:2010yg,Bern:2010ue}.
We begin by noting that a general massless $m$-point $L$-loop gauge
theory amplitude ${\cal A}_m^{(L)}$ in $d$ space-time dimensions can
be written as
\begin{equation}
{\cal A}_m^{(L)}=i^Lg^{m-2+2L}\sum_{i\in\Gamma}\int\prod_{l=1}^L\frac{d^dp_l}
{(2\pi)^d}\frac{1}{S_i}\frac{n_i\,c_i}{\prod_{\alpha_i}p_{\alpha_i}^2},
\label{ampform}
\end{equation}
where $g$ is the coupling constant. Here the sum is over the
complete set of graphs involving triple vertices, consistent
with the given loop order and number of external particles, and $S_i$
a symmetry factor for each graph $i$ (the dimension of its
automorphism group). The denominator contains all relevant propagator
momenta, and $n_i$ is the kinematic numerator associated with each
graph. Finally, $c_i$ is the colour factor of each graph, obtained by
dressing each triple vertex with structure constants. The
restriction to cubic diagrams in eq.~(\ref{ampform}) uses the fact
that four-gluon vertices can always be replaced with sums of products
of three-gluon vertices. Furthermore, this representation is not
unique - one may shuffle contributions between terms in the sum using
(generalised) gauge transformations. \\

As explained in detail in~\cite{Bern:2008qj}, one may classify cubic
diagrams into overlapping sets of three, whose colour factors are
related by Jacobi identities, originating from the Lie algebra of the
gauge group. BCJ duality is then the statement that, for each such
Jacobi identity, the kinematic numerators (as functions of the set of
independent external and loop momenta) can be chosen to obey the same
relation. We take this as an indication that some kinematic symmetry
algebra underlying the numerators exists. It is difficult to relate
this kinematic algebra, in general, to a symmetry of the Lagrangian of
the theory; a specific obstruction is the presence of the four-gluon
operator in the Lagrangian.
However,
in some cases (such as the self-dual sector of Yang-Mills theory), BCJ
duality can be made exact at the Lagrangian level, and the kinematic
symmetry interpreted~\cite{Monteiro:2011pc}. We return to this
point in the following section. String theory amplitudes, in particular using the heterotic string \cite{Tye:2010dd} and the pure spinor formalism \cite{Mafra:2011kj,Mafra:2014oia}, have also provided important insights into this algebra. \\

BCJ duality is not manifest for arbitrary choices of the numerators in
eq.~(\ref{ampform}). Rather, the BCJ conjecture states that at least
one choice of numerators exists such that the duality is
manifest. Once this has been achieved, the {\it double copy} states
that
\begin{equation}
{\cal M}_m^{(L)}=i^{L+1}\left(\frac{\kappa}{2}\right)^{m-2+2L}
\sum_{i\in\Gamma}\int\prod_{l=1}^L\frac{d^dp_l}
{(2\pi)^d}\frac{1}{S_i}\frac{n_i\,\tilde{n}_i}{\prod_{\alpha_i}p_{\alpha_i}^2}
\label{ampform2}
\end{equation}
is an $m$-point, $L$-loop gravity amplitude, where the graviton has been defined via
\begin{equation}
g_{\mu\nu}=\eta_{\mu\nu}+\kappa h_{\mu\nu}.
\label{hdef}
\end{equation}
Equation~(\ref{ampform2}) has been obtained from eq.~(\ref{ampform})
by replacing the Yang-Mills coupling with its gravitational
counterpart:
\begin{equation}
g\rightarrow\frac{\kappa}{2},\quad \kappa^2=16\pi G_N,
\label{gsrep}
\end{equation}
where $G_N$ is Newton's constant. Furthermore, the colour factors
$\{c_i\}$ in the gauge theory have been replaced with a second set of
kinematic numerators $\{\tilde{n}_i\}$, that may or may not come from the
same gauge theory. The gravity theory thus obtained depends on the
choice of the two gauge theories (e.g. on the amount of
supersymmetry). If both gauge theories are the pure
non-supersymmetric Yang-Mills theory, the gravity theory is General
Relativity coupled to a 2-form field (equivalent to an axion in four dimensions) and a dilaton~\footnote{For recent work
  regarding the construction of pure gravity (no additional fields) from a double copy procedure,
  see
  ref.~\cite{Johansson:2014zca}.}~\cite{Bern:2010yg,Bern:2010ue}. \\

Similarly, one may start with eq.~(\ref{ampform}) and replace the
kinematic numerators $\{n_i\}$ with a second set of colour factors
$\tilde{c}_i$:
\begin{equation}
{\cal T}_m^{(L)}=i^Ly^{m-2+2L}\sum_{i\in\Gamma}\int\prod_{l=1}^L\frac{d^Dp_l}
{(2\pi)^D}\frac{1}{S_i}\frac{\tilde{c}_i\,c_i}{\prod_{\alpha_i}p_{\alpha_i}^2},
\label{ampformscal}
\end{equation}
where $y$ is the appropriate coupling constant. The particle content
of this theory is a set of scalar fields $\phi^{aa'}$, which transform
in the adjoint representation of two Lie algebras. This is an example
of a {\it biadjoint} scalar theory \cite{Bern:1999bx,Du:2011js,BjerrumBohr:2012mg,Cachazo:2013iea}, as mentioned in the
introduction. The equation of motion of such a theory is explicitly
given by
\begin{equation}
\partial^2 \phi^{a a'} - y f^{abc} f^{a'b'c'} \phi^{bb'} \phi^{cc'} = 0,
\label{biadjointeom}
\end{equation}
where the second term arises from a cubic interaction involving both
sets of structure constants.\\

As is clear in the above discussion, the definitions of BCJ
duality and the double / zeroth copies are intrinsically
perturbative. Nevertheless, we will see the ideas of this section
(such as the replacement of coupling constants and colour information
by kinematics) throughout the paper. We will also use the Minkowski metric
$\textrm{diag}(-,+,+,+,\ldots)$ throughout, unless otherwise stated.

\section{Kerr-Schild coordinates and the double copy}
\label{sec:Kerr-Schild}

In this section, we examine a particular choice of coordinates in
gravity theories, namely {\it Kerr-Schild (KS) coordinates}, that will
be crucial for what follows. These coordinates are applicable to a
specific class of solutions of the Einstein equations, namely
Kerr-Schild solutions (for a review, see
e.g.~\cite{Stephani:2003tm}). These have the property that the
spacetime metric $g_{\mu \nu}$ may be written in the
form~\footnote{Throughout the paper we will refer to $h_{\mu\nu}$ (the
  deviation from the Minkowski metric) as the graviton, despite the
  fact that we are working in the non-linear regime.}
\begin{align}
\label{eq:ksdef}
g_{\mu \nu}&=\eta_{\mu\nu} +h_{\mu\nu}\notag\\
&\equiv \eta_{\mu \nu} + k_\mu k_\nu \phi
\end{align}
where $\phi$ is a scalar function, $\eta_{\mu \nu}$ is the Minkowski
metric and the (co)vector field $k_\mu$ has the property that it is null
with respect to both the Minkowski and full metric:
\begin{equation}
\eta^{\mu \nu} k_\mu k_\nu = 0 = g^{\mu \nu} k_\mu k_\nu.
\end{equation}
Therefore, the inverse metric is simply
\begin{equation}
g^{\mu \nu} = \eta^{\mu \nu} - \phi k^\mu k^\nu,
\end{equation}
where we raise the index on $k$ using the Minkowski metric. \\

To understand the dynamics of Kerr-Schild metrics, we turn to the
Einstein equations. Recall that the Einstein tensor is
\begin{equation}
G_{\mu\nu}=R_{\mu\nu}-\frac{1}{2}g_{\mu\nu}R,
\label{Einstein2}
\end{equation}
using the conventional notation for the Ricci tensor $R_{\mu\nu}$ and
scalar $R={R^\mu}_\mu$. In terms of the function $\phi$ and vector
$k_\mu$ introduced in Eq.~\eqref{eq:ksdef}, one
has
\begin{align}
R^\mu{}_\nu&=\frac{1}{2} \left(\partial^\mu \partial_\alpha \left(\phi k^\alpha k_\nu\right)+\partial_\nu \partial^\alpha \left(\phi k_\alpha k^\mu\right)-\partial^2 \left(\phi k^\mu k_\nu\right)\right);\notag\\
R&=\partial_\mu\partial_\nu\left(\phi k^\mu k^\nu\right),
\label{Rdefs}
\end{align}
where, as usual, $R^\mu{}_\nu = g^{\mu \lambda} R_{\lambda \nu}$ but
we have defined $\partial^\mu = \eta^{\mu \nu} \partial_\mu$. This
mixed convention is useful, because it is only with this particular
combination of indices that the Ricci tensor has the remarkable
property that it is linear in $\phi$ (and indeed in $h_{\mu\nu}$).\\

A simplification occurs for the stationary case in which all time
derivatives vanish. Without loss of generality, we may also set
$k^0=1$, with all dynamics in the zeroth component contained in the
function $\phi$. The stationary nature of the solution means that the
Ricci tensor can be simplified for each of the components, and one
finds (using $\partial_i=\partial^i$, $k_i=k^i$)
\begin{align}
R^{0}{}_{0}&=\frac{1}{2}\nabla^2 \phi ;\label{R00val}\\
R^{i}{}_0&=-\frac{1}{2}\partial_j \left[\partial^i\left(\phi k^j\right)-\partial^j\left(\phi k^i\right)\right];\label{R0ival}\\
R^i{}_j&=\frac{1}{2}\partial_l\left[\partial^i\left(\phi k^l k_j\right)+\partial_j\left(\phi k^l k^i\right)-\partial^l\left(\phi k^i k_j\right)\right];\label{Rijval}\\
R&=\partial_i\partial_j\left(\phi k^i k^j\right),\label{Rval}
\end{align}
where Latin indices run over the spacelike components. \\

Let us now interpret these equations in the spirit of the double
copy. To that end, we define a vector field $A_\mu = \phi k_\mu$, with
associated Abelian field strength $F_{\mu \nu} = \partial_\mu A_\nu -
\partial_\nu A_\mu$. We will refer to this as a Kerr-Schild
ansatz. More generally, one can consider a non-Abelian gauge field
$A^a_\mu$ which, in this Kerr-Schild ansatz, can be written as
$A^a_\mu = k_\mu \phi^a$.  The vacuum Einstein equations $R_{\mu \nu}
= 0$ imply, in the stationary case, that the gauge field satisfies the
(Abelian) Maxwell equations
\begin{equation}
\partial_\mu F^{\mu \nu} = \partial_\mu (\partial^\mu(\phi k^\nu)-\partial^\nu(\phi k^\mu)) =0,
\end{equation}
whose components are related to \eqref{R00val} and \eqref{R0ival}. It may seem surprising that the gauge field satisfies the Abelian
equations. However, this reflects the linear structure of the Einstein
equations in the Kerr-Schild coordinate system. \\

Going further, we may also interpret $\phi$ in the spirit of the
zeroth copy. The Kerr-Schild ansatz for the gauge field $A_\mu$ is
obtained by removing a factor of $k_\nu$ from the (non-perturbative) graviton
$h_{\mu\nu}$. Repeating this, we find that the Kerr-Schild scalar
function $\phi$ is the field that survives upon taking the zeroth
copy. This field then satisfies the equation of motion
\begin{equation}
\nabla^2 \phi = 0.
\label{phiabel}
\end{equation}
Thus, we see that eq.~(\ref{phiabel}) is an Abelianised version of the
biadjoint field equation of eq.~(\ref{biadjointeom}). It is important
to note that the scalar field $\phi$ plays a role analogous to the
propagators in the amplitudes story. It is present, and unchanged, in the
scalar, gauge and gravitational cases. Meanwhile, the gauge field
$A_\mu$ is obtained by multiplying the scalar by a single copy of a
vector field $k_\mu$ while the gravitational field is obtained by
multiplying the scalar by two copies of the vector
$k_\mu$. Furthermore, eq.~(\ref{phiabel}) gives us a direct physical
interpretation of the scalar field $\phi$. In the zeroth copy theory,
considering the general case in which a source term is also present,
the field $\phi$ will be the Green's function (scalar propagator)
integrated over the source. This is the same idea as in the double
copy of refs.~\cite{Bern:2010yg,Bern:2010ue}, reviewed here in
section~\ref{sec:amplitude}. In that case, one leaves propagator
denominators intact. They correspond to the scalar propagators that
one would obtain after taking a zeroth copy. \\

Let us make a few more remarks on the Kerr-Schild ansatz and the
double copy. In general, the squaring map from gauge theory to gravity
includes also a dilaton and a 2-form field. This follows from
degree-of-freedom counting: two states in gauge theory squares into
four states. Our double copy, using the Kerr-Schild ansatz for both
gauge theory and gravity, can be expressed as $A_\mu = k_\mu \phi,
h_{\mu\nu} = k_\mu k_\nu \phi$. This is manifestly symmetric, so the
2-form field is not present. In addition, since $k^2 = 0$ for the KS ansatz,
the trace of the graviton is manifestly zero, removing the
dilaton. Thus, the double copy map in this case is simply a map
between gauge theory and Einstein gravity. \\

In this section, we have considered stationary Kerr-Schild solutions. We will see explicit examples in what follows. First, it is instructive to provide further
motivation for relating this (Kerr-Schild) structure to the standard
BCJ double copy of refs.~\cite{Bern:2010yg,Bern:2010ue}. To this end,
one may examine the relationship between scattering amplitudes in
self-dual Yang-Mills theory and gravity~\cite{Monteiro:2011pc}, which admits a neat reinterpretation using the
Kerr-Schild language, albeit in momentum space. This is the subject of
the following section.

\subsection{Kerr-Schild-like approach to self-dual solutions}
\label{sec:selfdual}

In the previous section, we reviewed the properties of Kerr-Schild
solutions in General Relativity. These are a special class of
solutions whose general form is that of eq.~(\ref{eq:ksdef}), where
$k^\mu$ is defined as a function of the spacetime coordinates i.e. in
position space. From the point of view of particle scattering, it is
natural to apply a similar ansatz in momentum space, which corresponds
to the vector $k^\mu$ becoming a differential operator in position
space, $k^\mu\rightarrow\kh^\mu$. To this end, let us suppose that we
can write the metric as
\begin{align}
g_{\mu \nu} &= \eta_{\mu \nu} + \kappa h_{\mu\nu}\\
 &= \eta_{\mu \nu} + \kappa \kh_\mu \kh_\nu(\phi)
\label{metricKS}
\end{align}
where to be general we take $\kh_\mu$ to be an arbitrary linear
differential operator. We assume that this operator $\kh$ commutes with
itself, $[\kh_\mu, \kh_\nu] = 0$, in order that the metric be
symmetric. For example, we may take $\kh_\mu = \alpha_\mu^\nu
\partial_\nu$ where $\alpha$ is a constant matrix. We also restrict
our attention to double copies with no dilaton field. We therefore
wish to restrict to trace-free $h_{\mu\nu}$. Thus, we assume that $\kh^2
= 0$ in the sense that $\eta^{\mu \nu} \kh_\mu
\kh_\nu(\phi)=0$. Furthermore, we take
$\kh_\mu(\phi)\eta^{\mu\nu}\kh_\nu(\psi)=0$. Hence, the inverse metric is
\begin{equation}
g^{\mu \nu} = \eta^{\mu \nu} - \kappa \kh^\mu \kh^\nu(\phi).
\end{equation}
The solutions that underlie the double-copy in the self-dual sector (we are now considering four space-time dimensions) are precisely of this form.
Using light-cone coordinates 
\begin{equation}
u = t - z, \quad v = t + z, \quad w = x + i y\quad \bar w = x -i y,
\label{lightconecoords}
\end{equation}
such that the Minkowski line element becomes
\begin{equation}
ds^2=-du\,dv+dw\,d\bar{w},
\label{ds2lc}
\end{equation}
we can express the relevant $\kh^\mu$ for self-dual theory as
\begin{equation}
\kh_u=0,\quad \kh_v=\frac{1}{4}\partial_w,\quad \kh_w=0,\quad \kh_{\bar{w}}=\frac{1}{4}\partial_u.
\label{klc}
\end{equation}
Note that $\kh\cdot\partial\equiv 0$. The Christoffel symbols are
\begin{equation}
\label{eq:sdChristoffel}
\Gamma^\rho{}_{\mu \nu} = \frac {\kappa}{2} \left( \partial_\mu \kh^\rho \kh_\nu \phi + \partial_\nu \kh^\rho \kh_\mu \phi - \partial^\rho \kh_\mu \kh_\nu \phi + \kappa (\kh^\rho \kh^\sigma \phi) (\partial_\sigma \kh_\mu \kh_\nu \phi) \right),
\end{equation}
where we have raised the index on the partial derivatives using the
Minkowski metric. Using
Eq.~\eqref{eq:sdChristoffel}, we find that the vacuum Einstein
equations are 
\begin{equation}
\label{eq:sdEinstein}
R_{\mu\nu} = \frac {\kappa}{2}\left[ - \kh_\mu \kh_\nu \partial^2 \phi + \kappa (\kh_\mu \kh_\nu \partial_\rho \partial_\sigma \phi) (\kh^\rho \kh^\sigma \phi) - \kappa (\kh_\mu \kh_\rho \partial^\sigma \phi) (\kh_\nu \kh_\sigma \partial^\rho \phi)\right] = 0.
\end{equation}
This differs from the usual Kerr-Schild case because the Ricci tensor
is no longer linear in $\phi$; we recover linearity if we allow
$\kh_\mu$ to be an ordinary vector (rather than a linear operator). \\

The
Einstein equation~\eqref{eq:sdEinstein} is equivalent to a single
scalar equation
\begin{equation}
\partial^2 \phi - \frac{\kappa}{2}  (\kh^\mu\kh^\nu \phi)(\partial_\mu \partial_\nu \phi) = 0.  
\label{GravKS}
\end{equation}
Expanding this using eq.~\eqref{klc}, one finds that this is none other than the Plebanski
equation for self-dual gravity. The Plebanski equation is
known to provide an explicit example of a classical double copy, where the gauge theory
side is given by a particular form of the self-dual Yang-Mills equation~\cite{Monteiro:2011pc}. The next order of business is therefore to examine the Yang-Mills side of the correspondence. \\ 

We can perform a similar manipulation for self-dual Yang-Mills
theory. In this case, we assume that the gauge field has the
``Kerr-Schild" form
\begin{equation}
A^a_\mu = \kh_\mu \phi^a,
\label{AmuKS}
\end{equation}
where $\phi^a$ are Lie-algebra-valued scalars, and the 4-vector $\kh$ is the linear
differential operator appearing in eq.~(\ref{klc}). It is
straightforward to calculate the Yang-Mills equations in this gauge;
the result is
\begin{equation}
\kh_\nu \partial^2 \phi^a + 2 g f^{abc} (\kh^\mu \phi^b) (\kh_\nu\partial_\mu \phi^c) = 0.
\label{YMKS}
\end{equation}
Multiplying by $T^a$ and expanding using the metric in eq.~\eqref{ds2lc}
and eq.~\eqref{klc} yields
\begin{equation}
\kh_\nu\left(\partial^2\Phi+ig\left[\partial_w\Phi,\partial_u\Phi\right]\right)=0,
\label{YMKS2}
\end{equation}
where $\Phi=\phi^aT^a$. This equation is equivalent to the standard
self-dual YM equation, in the form which makes the double copy
manifest (see ref.~\cite{Monteiro:2011pc} for a detailed review). We
therefore see that this self-dual double copy arises from a
momentum-space Kerr-Schild description. This is appealing in that the
Kerr-Schild ans\"{a}tze for the gauge field and graviton,
eqs.~\eqref{metricKS} and~\eqref{AmuKS}, are manifest double copies of
each other. \\

In this section, we have seen that the self-dual scattering amplitudes
of ref.~\cite{Monteiro:2011pc} can be expressed using a momentum-space
generalisation of Kerr-Schild coordinates. We are now further
encouraged in our identification of the stationary Kerr-Schild double
copy structure of section~\ref{sec:Kerr-Schild} with the standard BCJ
double copy. Let us move on to study some examples.

\section{Stationary Kerr-Schild solutions}
\label{sec:Kerr2}
We have already seen in section~\ref{sec:Kerr-Schild} that there is a special
class of gravitational Kerr-Schild solutions, which map to solutions of
the Abelian Yang-Mills equations. Given the comments in the previous
section, we can make this more formal as follows: let
eq.~(\ref{eq:ksdef}) be a stationary solution of the Einstein
equations ($\partial_0\phi=\partial_0 k^\mu=0$), and choose $k^0=1$. Then
\begin{equation}
A^\mu_a=c_a \,\phi\,  k^\mu
\label{Amudef3}
\end{equation}
is a solution of the Yang-Mills equations, for an arbitrary choice of
constants $c_a$ (since this ansatz linearises the Yang-Mills
equations). This constitutes a large general class of solutions that
can be identified between gauge and gravity theories. We can then
refer to the gauge theory solution as a {\it single copy}, or ``square
root'', of the gravity solution. Let us begin with a familiar example.

\subsection{The Schwarzschild black hole}
\label{sec:Schwarzschild}

The Schwarzschild black hole is the most general spherically symmetric
solution of the vacuum Einstein equations. It must be static and
asymptotically flat, by Birkhoff's theorem (see
e.g.~\cite{Stephani:1982ac}). One can source this with a pointlike
mass $M$, via the energy-momentum tensor
\begin{equation}
T^{\mu\nu}=Mv^\mu v^\nu\delta^{(3)}({\mathbf x}),
\label{BHsource}
\end{equation}
where $v^\mu=(1,0,0,0)$ is a vector pointing purely in the time
direction. The solution of the Einstein equations
\begin{equation}
G_{\mu\nu}=\frac{\kappa^2}{2}T_{\mu\nu}
\label{Einstein}
\end{equation}
for the metric $g_{\mu\nu}$ depends on the choice of gauge - in this
case a choice of coordinate system. It is well-known, however, that a
Kerr-Schild form exists, in which the exterior Schwarzschild metric
has the form
\begin{equation}
g_{\mu\nu}=\eta_{\mu\nu}+\frac{2GM}{r}k_\mu k_\nu,
\label{kerrschild}
\end{equation}
where $G$ is Newton's constant, and 
\begin{equation}
k^\mu=\left(1,\frac{x^i}{r}\right),\quad r^2=x^i x_i,\quad i=1\ldots3.
\label{kdef}
\end{equation}
Comparing eq.~(\ref{kerrschild}) with eq.~(\ref{hdef}) and using the
relation
\begin{equation}
\kappa^2=16\pi G,
\label{kappadef}
\end{equation}
one finds, using a convenient normalisation, that the (non-perturbative) graviton for the
exterior Schwarzschild solution is given by
\begin{equation}
h_{\mu\nu}=\frac{\kappa}{2}\phi k_\mu k_\nu,\quad \phi=\frac{M}{4\pi r}.
\label{hBH}
\end{equation}
According to the results of section~\ref{sec:Kerr-Schild}, one may
take a single copy of this solution:
\begin{equation}
A^\mu=\frac{g c_aT^a}{4\pi r}\left(1,\frac{\mathbf x}{r}\right)\equiv
\frac{g c_a T^a}{4\pi r}k_\mu,
\label{singlecopy}
\end{equation}
where $k_\mu$ is defined as in eq.~(\ref{kdef}). We have obtained this
from eq.~(\ref{hBH}) via the replacements
\begin{equation}
\frac{\kappa}{2}\rightarrow g,\quad M\rightarrow c_a T^a,\quad k_\mu k_\nu
\rightarrow k_\mu,\quad \frac{1}{4\pi r}\rightarrow\frac{1}{4\pi r} .
\label{replace2}
\end{equation}
This all makes very good sense from a double copy point of view. The
first replacement is the usual identification of coupling constants in
the two theories. The second replacement replaces a charge in the
gravity theory (a mass) with a corresponding colour charge. Indeed,
substituting eq.~(\ref{singlecopy}) into the Abelian Maxwell
equations, one finds
\begin{equation}
\partial_\mu F^{\mu\nu}=j^\nu,
\label{abmax}
\end{equation}
where the current
\begin{equation}
j^\nu=-g (c_a T^a) v^\nu\delta^{(3)}({\mathbf x})
\label{jmuBH}
\end{equation}
corresponds to a static colour source located at the origin, with
4-velocity $v^\mu=(1,\vec{0})$. It is interesting that this happens:
the double copy is a statement about the gauge field and
graviton. Here we have inserted extra degrees of freedom into the
theory, for the purpose of sourcing the gauge fields. The single copy,
applied to the fields, has correctly identified the required source in
the gauge theory. The final replacement in eq.~(\ref{replace2})
corresponds to a scalar propagator that remains unchanged on the
gravity side, in line with our previous comments.\\

Let us now physically interpret the gauge solution of
eq.~(\ref{singlecopy}). Given that this is a solution of the Abelian
Maxwell equations, we can perform a gauge transformation according to
\begin{equation}
A^a_\mu\rightarrow A^a_\mu+\partial_\mu\chi^a(x).
\label{gaugetrans}
\end{equation}
Let us choose
\begin{equation}
\chi^a=-\frac{gc_a}{4\pi}\log\left(\frac{r}{r_0}\right),
\label{chidef}
\end{equation}
where $r_0$ is an arbitrary length scale to make the logarithm
dimensionless. In the new gauge, one has
\begin{equation}
A_\mu=\left(\frac{g c_a T^a}{4\pi r},0,0,0\right).
\label{Aadef}
\end{equation}
This is recognisable as the Coulomb solution for the superposition of
static colour charge that we have located at the origin, which is not
surprising. It is well-known in Maxwell electromagnetism, for example,
that the most general solution consistent with spherical symmetry is
the Coulomb solution plus a radiation field. A static
solution rules out the latter. Our results are also consistent with
ref.~\cite{Sikivie:1978sa}, which addresses the solution of the
Yang-Mills equations with an arbitrary static source. The authors
argue that one may gauge away the non-Abelian nature of the Yang-Mills
field, leaving the Abelian-like Coulomb solution.\\

We stress that the double copy between the Coulomb and Schwarzschild
solutions only manifests itself with a particular gauge choice. The
role of Kerr-Schild coordinates on the gravity side was crucial here.
Before moving on, it is worth pointing out that the Schwarzschild
double copy works also in higher dimensions, as it must do if our
double copy interpretation is correct. The $d$-dimensional
generalisation of the Schwarzschild black hole was first found by
Tangherlini~\cite{Tangherlini:1963bw}, and in Kerr-Schild coordinates
the metric is given by
\begin{equation}
g_{\mu\nu}=\eta_{\mu\nu}+\frac{\mu}{r^{d-3}}k_\mu k_\nu,
\label{schwarzschildN}
\end{equation}
where $k^\mu$ is a simple generalisation of eq.~(\ref{kdef}), such
that $i=1\ldots (d-1)$, and the parameter $\mu$ is related to the mass
$M$ via~\footnote{Note that in eq.~(\ref{Mdef}) we have absorbed a
  factor of $2/(d-2)$ into Newton's constant. This ensures that the
  relation $\kappa^2=16\pi G$ holds for arbitrary values of $d$.}
\begin{equation}
M=\frac{\Omega_{d-2}}{8\pi G}\mu,
\label{Mdef}
\end{equation}
with
\begin{equation}
\Omega_{d-2}=\frac{2\pi^{\frac{d-1}{2}}}{\Gamma\left(\frac{d-1}{2}\right)}
\label{omegadef}
\end{equation}
the area of a unit $(d-2)$-sphere. The Newtonian
potential obtained from the metric of eq.~(\ref{schwarzschildN}) is
\begin{equation}
\phi=\frac{4\pi}{\Omega_{d-2}}\frac{GM}{r}.
\label{phires}
\end{equation}
Taking the single copy of this result, one obtains
\begin{equation}
A^\mu=\frac{g T_a}{\Omega_{d-2}}k^\mu.
\label{AmuN}
\end{equation}
This can be obtained from the $d$-dimensional Coulomb solution
\begin{equation}
A^\mu=\left(\frac{g T_a}{\Omega_{d-2}r^{d-3}},\vec{0}\right)
\label{CoulombN}
\end{equation}
via a gauge transformation according to eq.~(\ref{gaugetrans}),
choosing
\begin{equation}
\chi=\int^r dr' A^0(r')=\frac{g T_a}{(4-d)\Omega_{d-2}}r^{4-d},
\label{chiN}
\end{equation}
valid for $d>4$. The uniqueness of the Coulomb-like solution can again
be understood from an electromagnetic analogue of Birkhoff's theorem,
which fixes the Tangherlini solution in GR.\\

\subsection{The Kerr black hole}
\label{sec:Kerr}

In the previous subsection, we have seen how the Kerr-Schild story of
section~\ref{sec:Kerr-Schild} can be used to obtain a single copy of
the Schwarzschild solution, namely a Coulomb solution. We saw also
that the sources for the two solutions had a double copy structure,
with a static colour charge on the gauge side becoming a static mass
in the gravity theory. In this section we study another example,
namely the Kerr (uncharged, rotating) black hole. In Kerr-Schild
coordinates, the graviton field is
\begin{equation}
h_{\mu\nu}=\phi(r)k_\mu k_\nu,
\label{kerr}
\end{equation}
where
\begin{equation}
\phi(r)=\frac{2MGr^3}{r^4+a^2z^2}
\label{fdef}
\end{equation}
and 
\begin{equation}
k_\mu=\left(1,\frac{rx+ay}{r^2+a^2},\frac{ry-ax}{r^2+a^2},\frac{z}{r}\right).
\label{ldef}
\end{equation}
Note that $r$ is no longer simply the modulus of the vector
$(x,y,z)$. It is instead defined implicitly via the equation
\begin{equation}
\frac{x^2+y^2}{r^2+a^2}+\frac{z^2}{r^2}=1
\label{rdef}
\end{equation}
except for the region $\{x^2+y^2\leq a^2, z=0\}$ (i.e. a disc of
radius $a$ about the origin in the $(x,y)$ plane), where $r=0$. 
\\

Following the Kerr-Schild single copy procedure as we did for the
Schwarzschild solution, one may construct the gauge
field
\begin{equation}
A^a_\mu=\frac{g}{4\pi}\phi(r)c_ak_\mu,
\label{Amukerr}
\end{equation}
where again this is a solution to the Abelian Maxwell equations in the
vacuum region described above. The rotation introduces a magnetic component to the Maxwell field.\\

We may interpret the gauge and gravity solutions further by
determining the sources that create them. Given that we are only
concerned with the vacuum solution for the Kerr metric, this means
that we should look for the minimal possible source that will generate
this. The source is a disk whose mass distribution exhibits a ring singularity
at $x^2+y^2=a^2$ (which generates the well-known curvature
singularity of the metric there). This was first explored by
Israel~\cite{Israel:1970kp}; we have also found ref.~\cite{Balasin:1993kf} useful.  \\

One may introduce the spheroidal coordinates
\begin{equation}
x=\sqrt{r^2+a^2}\sin\theta\cos\phi,\quad y=\sqrt{r^2+a^2}\sin\theta\sin\phi,\quad z=r\cos\theta,
\label{spheroids}
\end{equation}
where $r$, $\theta$ and $\phi$ play the role of a radial, polar
angular and azimuthal angular coordinate respectively. Surfaces of
constant $r=R$ are ellipsoids, such that $R\rightarrow0$ converges to
the disk of radius $a$ in the $(x,y)$ plane. In these coordinates, the
energy-momentum tensor one finds for the Kerr metric
is~\cite{Israel:1970kp}~\footnote{Note that our definition of $\sigma$
  in eq.~(\ref{TKerr2}) differs from that of~\cite{Israel:1970kp} by a
  factor of two, due to us having normalized our surface integral as
  being over both the upper and lower surfaces of the disk.}
\begin{equation}
T^{\mu\nu}=\sigma\left(w^\mu w^\nu+\zeta^\mu\zeta^\nu\right),\quad \sigma=-\frac{M}{8\pi^2 a\cos\theta}\delta(z)\Theta(a-\rho),
\label{TKerr2}
\end{equation}
where we have introduced the radial and spacelike 4-vectors (in the
spheroidal coordinate system $(t,r,\phi,\theta)$)
\begin{equation}
w^\mu=\tan\theta\left(1,0,1/(a\sin^2\theta),0\right),\quad \zeta^\mu=\left(0,1/(a\cos\theta),0,0\right),
\label{wzetadef}
\end{equation}
and also defined
\begin{equation}
\rho=a\sin\theta.
\label{rhodef}
\end{equation}
As discussed in~\cite{Israel:1970kp}, this has the form of a negative
proper surface density rotating about the $z$-axis with superluminal
velocity, and balanced by a radial pressure. The former and latter
effects are the first and second terms in eq.~(\ref{TKerr2})
respectively. \\

We may repeat the above exercise for the square root of the Kerr
solution. Substituting the gauge field into the Abelian Maxwell
equations, one finds a source current
\begin{equation}
j^\mu=-\delta(z)\Theta(a-\rho)\frac{g(c_aT^a)}{4\pi}\frac{1}{a^2\cos\theta}\left(\sec^2\theta,0,\frac{\sec^2\theta}{a},0\right).
\label{jKerr}
\end{equation}
Here we have performed the coupling constant and mass replacements of
eq.~(\ref{replace2}). Introducing the vector~\cite{Israel:1970kp}
\begin{equation}
\xi^\mu=\left(1,0,\frac{1}{a},0\right), 
\label{xidef}
\end{equation}
one may rewrite eq.~(\ref{jKerr}) as 
\begin{equation}
j^\mu=q \xi^\mu,\quad q=-\delta(z)\Theta(a-\rho) \frac{g(c_aT^a)}{4\pi a^2}\sec^3\theta.
\label{jKerr2}
\end{equation}
This has the form of a distribution of colour charge rotating about
the $z$-axis. \\

Some additional comments are in order regarding the source terms we
have found above. Firstly, it is not obviously the case that the
energy-momentum tensor of eq.~(\ref{TKerr2}) is a ``double copy'' of
the current of eq.~(\ref{jKerr2}). Perhaps the correct way to view this,
however, is as follows. One may write eq.~(\ref{TKerr2}) as 
\begin{equation}
T^{\mu\nu}=\delta(z)\Theta(a-\rho)\left(-\frac{M\sec^3\theta}{8\pi a^2}\right)\left[\xi^\mu\xi^\nu -\cos^2\theta\, \tilde{\eta}^{\mu\nu}\right]
\label{TKerr3}
\end{equation}
where, in the Cartesian coordinates $(t,x,y,z)$, $\tilde{\eta}^{\mu\nu}=\textrm{diag}(-1,1,1,0)$.
The first term of eq.~(\ref{TKerr3}) consists of a pure double copy of
the gauge theory solution, where the rotating charge has been replaced
with a similar rotating mass distribution. However, this would not be stable
in the gravity theory, and thus would not lead to a static solution for
the metric. The additional terms in eq.~(\ref{TKerr3}) thus supply the
radial pressure that is needed to stabilise the disk.\\

The Kerr black hole has a higher-dimensional extension, known as the
Myers-Perry black hole \cite{Myers:1986un}. A major difference is
that, in $d$ space-time dimensions, there are $(d-1)/2$ independent
rotation planes if $d$ is odd, and $(d-2)/2$ if $d$ is even; this is
the dimension of the Cartan subgroup of $SO(d-1)$. So multiple angular
momenta are allowed, one per rotation plane, which makes the problem
challenging. This challenge is alleviated by the fact that Myers-Perry
black holes are also Kerr-Schild solutions. We can take the graviton
field \eqref{kerr}, and write the general solution as
\begin{equation}
    \phi(r) =
    \begin{cases}
     \displaystyle{ \frac{\mu r^2}{\Pi F}} \quad & \text{if } d \text{ is odd},\\
     \phantom{a} & \phantom{a} \\
     \displaystyle{\frac{\mu r}{\Pi F}} \quad & \text{if } d \text{ is even},
    \end{cases}
\end{equation}
and
\begin{equation}
    k_\mu dx^\mu =
    \begin{cases}
     \displaystyle{ dt + \sum_{i=1}^{(d-1)/2} \frac{r(x^idx^i+y^idy^i)+a_i(x^idy^i-y^idx^i)}{r^2+a_i^2}} \qquad & \text{if } d \text{ is odd},\\
     \phantom{a} & \phantom{a} \\
     \displaystyle{ dt + \frac{zdz}{r} + \sum_{i=1}^{(d-2)/2} \frac{r(x^idx^i+y^idy^i)+a_i(x^idy^i-y^idx^i)}{r^2+a_i^2}} \qquad & \text{if } d \text{ is even}.
    \end{cases}
\end{equation}
For each rotation plane, there is a rotation parameter $a_i$, and a pair of coordinates $(x^i,y^i)$. We have used the functions
\begin{equation}
\Pi=\prod_{i=1}^{(d-2)/2} (r^2+a_i^2), \qquad F = 1- \sum_{i=1}^{(d-1)/2} \frac{a_i^2({x^i}^2+{y^i}^2)}{(r^2+a_i^2)^2}.
\end{equation}
Finally, the radial variable $r$ is defined via
\begin{equation}
 \sum_{i=1}^{(d-1)/2} \frac{({x^i}^2+{y^i}^2)}{(r^2+a_i^2)^2} =0 \quad \text{if } d \text{ is odd}, \qquad
 \frac{z^2}{r^2}+\sum_{i=1}^{(d-2)/2} \frac{({x^i}^2+{y^i}^2)}{(r^2+a_i^2)^2} =0 \quad \text{if } d \text{ is even}.
\end{equation}
The Myers-Perry black holes provide a straightforward extension of our discussion on the Kerr black hole. They allow for solutions to Maxwell's equations in higher dimensions based on the Kerr-Schild ansatz, $A_\mu=\phi \,k_\mu$. \\

We point out that not all vacuum asymptotically flat black holes in higher dimensions are of Kerr-Schild type. See \cite{Emparan:2008eg} for a review. Such solutions may have a variety of horizon topologies and are much harder to construct, the simplest example being the black ring (horizon topology $S^1\times S^2$ in five dimensions). It would be interesting to analyse their gauge theory counterparts. One way to proceed would be the following. In the case of five dimensions, there is a solution-generating technique for vacuum black holes because, after symmetry considerations, the Einstein equations reduce to a two-dimensional integrable system. If an extension of this technique to gauge theory exists, a map between the two types of solutions may be constructed. This would provide non-Kerr-Schild examples of the double copy.

\subsection{Black branes}
\label{sec:branes}

One other type of vacuum solution worth mentioning is that of black branes. These are black holes with horizons that extend in extra spatial dimensions. Since the direct product of two Ricci-flat manifolds is also a Ricci-flat manifold, solutions with extended horizons are trivially constructed from lower-dimensional black holes. Consider the metric of a ($d+m$)-dimensional black brane which uniformly extends a $d$-dimensional black hole with metric $g_{\mu\nu}$ along $\mathbb{R}^m$ ($\hat{\alpha}=1\ldots m$),
\begin{equation}
\label{blackbrane}
ds_{\text{brane}}^2 = g_{\mu\nu}dx^\mu dx^\nu + dz_{\hat{\alpha}} dz^{\hat{\alpha}}.
\end{equation}
The simplest example is the Schwarzschild string, $\textrm{Schwarz}_d \times \mathbb{R}$. The solutions discussed in the previous sections can therefore be trivially uplifted to new Kerr-Schild solutions,
\begin{equation}
\label{blackbraneKS}
g_{MN} = \eta_{MN} + \phi k_M k_N,
\end{equation}
where $M,N=1\ldots d+m$, and we have the extensions $\eta_{MN}=\textrm{diag}(-1,1,1,\ldots)$ and $k_M=\{k_\mu,0,0,\ldots\}$. The corresponding gauge theory solutions are then simply $A_M=\phi k_M$.\\

The reason why there is no such black string solution in four dimensions is that there is no vacuum asymptotically flat black hole in three dimensions. This case deserves some comments. Consider a ``cosmic string" which is a distribution of mass along the $z$ axis in four dimensions,
\begin{equation}
T^{MN}=\sigma v^M v^N \delta(x)\delta(y), \qquad v^M=(1,0,0,1).
\label{CSsource}
\end{equation}
This is an extension of the source \eqref{BHsource} along an extra dimension with coordinate $z$, where we have $d=3$ and $m=1$. Naively, the Kerr-Schild solution would be
\begin{equation}
\phi (r)= \sigma \log\frac{r}{r_0}, \qquad k^M=\left(1,\frac{x}{r},\frac{y}{r},0\right),
\end{equation}
with $r=\sqrt{x^2+y^2}$, where
$(\partial_x^2+\partial_y^2)\phi=0$. However, this solution is valid
only on the gauge theory side of the double copy, where one obtains
the standard electromagnetic field generated by a charged line. The
Kerr-Schild gravity solution constructed in this way does not satisfy
the vacuum Einstein equations, which is consistent with that fact
that, from eq.~(\ref{Rijval}), we see that the Einstein equations are
stronger than their gauge theory counterparts. In fact, it is well
known that the space-time solution corresponding to the source
\eqref{CSsource} is flat, except for a conical singularity on the $z$
axis (which has observational consequences). Despite the similarity
with previous cases, what happens here is that the Einstein equations
have no propagating degrees of freedom in three dimensions (the $z$
direction is a mere spectator), i.e. there is no graviton to be
obtained as the double copy of gluons.

\section{Time-dependent Kerr-Schild solutions}
\label{sec:waves}

We have now seen two specific examples in which stationary Kerr-Schild
solutions in gravity have well-defined single copies in Yang-Mills
theory. In this section, we briefly discuss well-known
time-dependent solutions.

\subsection{Plane wave solutions}
\label{sec:planewave}

Plane wave (pp-wave) solutions are arguably the simplest
time-dependent vacuum solutions in either gauge or gravity
theories. Unsurprisingly, gauge theory plane waves are related to gravitational pp-waves by a double copy prescription; this is manifestly the case for linearised waves (see e.g.~\cite{Siegel:1999ew}~\footnote{Linearised
  gravitational waves have recently been examined in the language of
  Maxwell's equations, in~\cite{Barnett:2014era}.}), and it extends to the non-perturbative case. Since pp-waves are Kerr-Schild solutions, they can be written as in equation~\eqref{eq:ksdef}. Using a pair of lightcone coordinates $x^\mu=(u,v,x^i)$, with $i=1\dots d-2$, we can express a pp-wave using
\begin{equation}
k_\mu dx^\mu =du = dz-dt, \qquad  \phi=\phi(u,x^i).
\label{ppwave}
\end{equation}
Then the Einstein equations are simply
\begin{equation}
\label{eq:ppwaveEOM}
\partial_i\partial^i \phi = 0.
\end{equation}
 Non-Abelian plane wave solutions also have the
form~\cite{Coleman:1977ps} 
\begin{equation}
A^a_\mu = k_\mu \phi^a(u,x^i), 
\label{Aaplane}
\end{equation}
where $\phi^a$ fulfills the ``propagator" equation~(\ref{eq:ppwaveEOM}). As
in the stationary cases considered previously, the Kerr-Schild
language makes the double copy explicit. Furthermore, the non-Abelian
solution is also a solution of the Abelian Yang-Mills equations. This is, of course, what happens on the gravity side: a
coordinate system has been chosen such that the expression for the
metric, expanded in $\kappa$, terminates at first order. The graviton field obtained is truly non-perturbative, but still admits a simple double copy interpretation.

\subsection{Shockwave solutions}
\label{sec:shockwave}

Perhaps the simplest case of a time-dependent field solution with a
source term is a shockwave. This corresponds to an infinitely boosted
particle, whose field (in either the gauge or gravity context) is
Lorentz contracted so that it lies in a flat plane transverse to the
particle direction. In gravity, shockwaves are described by the
Aichelburg-Sexl metric~\cite{Aichelburg:1970dh}. In four dimensions, the metric takes the pp-wave form \eqref{ppwave} with
\begin{equation}
\phi(u,x^i)=C\,\delta(u)\,\log|{\mathbf x}|.
\end{equation}
In their original
paper, they showed explicitly that one may obtain the shockwave metric
by a singular coordinate transformation of the Schwarzschild black
hole - namely, the coordinate transformation corresponding to an
infinite boost of the latter.\\

The perturbative relationship between the shockwave solutions in QCD
and gravity was recently discussed extensively
in~\cite{Saotome:2012vy}, which used Feynman diagram arguments in the
Regge limit (corresponding to the scattering of two highly boosted
particles) to perturbatively construct the shockwave solution in both
Yang-Mills theory and gravity, making clear the double copy
relationship~\footnote{See~\cite{Vera:2012ds,Johansson:2013nsa,Johansson:2013aca}
  for other recent work looking at the Regge limit from a double copy
  point of view.}. Furthermore, they found that the double copy was
insensitive to whether an abelian or non-abelian shockwave was used on
the gauge theory side. This is entirely consistent with our results
regarding the Schwarzschild metric. We have shown that this metric is a double copy of a point charge; boosting this in the gauge and gravity cases leads to the appropriate shockwaves in both theories. 

\section{Discussion}
\label{sec:conclusion}

In this paper, we have examined how certain classical gauge and gravity
solutions can be related by a double copy procedure, analagous to that
postulated to exist in perturbative quantum field
theory~\cite{Bern:2010yg,Bern:2010ue}.  In particular, we considered Kerr-Schild metrics in the gravity side of the
correspondence, and an associated Kerr-Schild ansatz for the gauge
field.  Kerr-Schild metrics have the property that the graviton
factorises into an outer product of two Lorentz vectors, dressed by an
overall scalar function. For the self-dual solutions, we used a slight
modification of the Kerr-Schild ansatz, and the results motivated us
to look for position-space examples.\\

We found an infinite class of Kerr-Schild double copies, namely those
for which the graviton and gauge field are stationary. This allowed us
to find single copies, or square roots, of well-known black hole
solutions (including the Schwarzschild and Kerr solutions, as well as
their higher-dimensional
generalisations). These single
copies were found to be solutions of the Abelian Yang-Mills (Maxwell)
equations, trivially embedded in the non-Abelian theory.  The sources
needed to generate these solutions also had an interesting double copy
interpretation, involving the replacement of colour charge by mass. \\

Our work leaves several questions unanswered. First, why should the
single copies of these stationary gravity solutions be Abelian? Perhaps we should not be surprised that the single copy gauge field is Abelian, in view of the remarkable fact that the Einstein equations for Kerr-Schild spacetimes are linear in $h_{\mu \nu}$. It
has been argued before for static sources (applicable to the
Schwarzschild case) that any non-Abelian component of the solution can
be gauged away~\cite{Sikivie:1978sa}. It may be simply that the
Abelian nature is forced upon the gauge theory by the use of
Kerr-Schild coordinates: for a double copy to a Kerr-Schild solution
to exist, it must be true that the gauge theory terminates at first
order in the coupling constant, forbidding the presence of
(non-Abelian) higher order corrections. Indeed, in the
Schwarzschild-Tangherlini examples, where the source is pointlike, it
is easy to see that a stationary, spherically symmetric solution of
the YM equations using the KS ansatz, which has the property that the
solution can be expanded perturbatively in the coupling, must be
Abelian. 
Another possibility is that there may exist non-Abelian solutions that
double copy to the same gravitational solution as an Abelian-like
gauge theory object. This is possible due to the fact that information
is lost when performing the double copy. An explicit example of this
is the analysis of ref.~\cite{Oxburgh:2012zr}, which showed that the
infrared singularities of QED and QCD both double copy to the infrared
singularities of GR. A related observation was made recently in
ref.~\cite{Bjerrum-Bohr:2013bxa}. It was found that gravitational
Compton scattering could be obtained as a double copy of Compton
scattering in either QCD or QED, depending on the choice of the
colour-ordered amplitudes involved in the double copy. Finally, we
point out that, at the level of the pure gauge theory amplitude, the
kinematic numerators (or the colour-ordered amplitudes) are the same
for photons and gluons. It is the colour dependence that leads to the
vanishing of the colour-dressed amplitude in the Abelian case.\\

A second question is whether or not the Kerr-Schild double copies are
genuine manifestations of the double copy as usually
defined~\cite{Bern:2010yg,Bern:2010ue}. There are a number of
arguments for why this is the case. Firstly, the Kerr-Schild picture
is analogous to the self-dual double copy, as discussed in
section~\ref{sec:selfdual}, as well as to other known cases of
classical double copies.  Secondly, the point source for the
Schwarzschild solution has an obvious single copy which
sources the Coulomb solution in gauge theory. This property is
reminiscent of the perturbative double copy of infrared singularities
discussed in ref.~\cite{Oxburgh:2012zr}. In fact, the Schwarzschild/Coulomb case has been studied recently in ref.~\cite{Neill:2013wsa}, where the field configurations were constructed from a perturbative approach. Finally, we interpreted the
Kerr-Schild function $\phi$ by taking a zeroth copy to a biadjoint
scalar theory. This, by definition, fixes that part of the gauge
theory solution that should not be squared upon performing the double
copy to gravity. In all cases we examined, the function $\phi$ was a
(scalar) propagator integrated over the distribution of charge, as
expected from the conventional double copy, in which denominators of
propagators are left intact. \\

In addition to stationary Kerr-Schild double copies, we mentioned
the case of pp-waves (shockwaves are a particular example), in which time-dependent solutions match under the double copy.
However, these solutions have a
simple time-dependence, that disappears in an infinite momentum
frame. We also found that the single copies of the
gravity solutions were Abelian-like in the gauge theory, which
may also be a consequence of the Kerr-Schild coordinates, as discussed
above. \\

We hope that this preliminary study provides a useful precursor for
further work. There are many avenues to be explored. It is fair to
say that we still lack a general understanding of which gauges should
be chosen in the gauge and gravity theories for the double copy to be
manifest. The recent proposal of ref.~\cite{Anastasiou:2014qba} has related the gauge symmetries on both sides using a convolution procedure for the double copy, but only at a linearised level. The Kerr-Schild solutions that we analysed here are a guide to further progress, in that more involved double copy procedures for non-perturbative solutions should explain the Kerr-Schild cases. It would be interesting then to analyse the consequences of a general double copy procedure. We expect that the theory of perturbations on backgrounds related by the double copy will itself exhibit double copy properties. It would
also be very interesting to examine quantum-corrected
solutions in the two theories, rather than purely classical
examples, perhaps following the lines of ref.~\cite{Bjerrum-Bohr:2013bxa}. Work in this regard is ongoing.

\section*{Acknowledgments}

We are very grateful to Simon Caron-Huot, Clifford Cheung, Christoph
Englert and David Miller for useful discussions on this topic. CDW is
supported by the Science and Technologies Facilities Council (STFC),
and is perennially thankful to the Higgs Centre for Theoretical
Physics for hospitality. RM is a Marie Curie Fellow, and also a Junior
Research Fellow at Linacre College, Oxford.

\bibliography{refs.bib}
\end{document}